\documentclass[
preprint,
 amsmath,amssymb,
 aps,
PRL,
longbibliography,
]{revtex4-1}

\usepackage{graphicx}
\usepackage{dcolumn}
\usepackage{bm}
\usepackage{color}
\usepackage{braket}
\usepackage{multirow}
\usepackage[colorlinks,linkcolor=blue, urlcolor=blue,citecolor=blue,bookmarks=fales]{hyperref}

\begin{document}

\title{Zeeman effects on Yu-Shiba-Rusinov states}

\author{T. Machida$^{1,2,\ast}$, Y. Nagai$^{3,4}$, and T. Hanaguri$^{1}$}

\affiliation{RIKEN Center for Emergent Matter Science, Wako, Saitama 351-0198, Japan}
\affiliation{Precursory Research for Embryonic Science and Technology (PRESTO), Japan Science and Technology Agency (JST), Tokyo 102-0076, Japan}
\affiliation{CCSE, Japan Atomic Energy Agency, 178-4-4, Wakashiba, Kashiwa, Chiba, 277-0871, Japan}
\affiliation{Mathematical Science Team, RIKEN Center for Advanced Intelligence Project (AIP), 1-4-1 Nihonbashi, Chuo-ku, Tokyo 103-0027, Japan}

\begin{abstract}
    When the exchange interaction between the impurity spin and the spins of itinerant quasiparticles is strong or weak enough, the ground states for a magnetic impurity in a superconductor are the screened or free spins, respectively.
    In both cases, the lowest excited state is a bound state within the superconducting gap, known as the Yu-Shiba-Rusinov (YSR) state.
    The YSR state is spatially localized, energetically isolated, and fully spin-polarized, leading to applications such as functional scanning probes.
    While any application demands identifying whether the impurity spin is screened or free, a suitable experimental technique has been elusive.
    Here we demonstrate an unambiguous way to determine the impurity ground state using the Zeeman effect.
    We performed ultra-low temperature scanning tunneling spectroscopy of junctions formed between a Cu(111) surface and superconducting Nb tips decorated by single magnetic Fe atoms.
    Depending on the condition of the Fe adsorbate, the YSR peak in the spectrum either splits or shifts in a magnetic field, signifying that the Fe spin is screened or free, respectively.
    Our observations provide renewed insights into the competition between magnetism and superconductivity and constitute a basis for the applications of the YSR state.
\end{abstract}

\maketitle

An antiferromagnetic exchange interaction $J$ between the spins of the impurity and itinerant quasiparticles brings about Kondo screening that competes with the superconducting pairing interaction.
If $|J|$ is large enough, the Kondo temperature $T_{\rm K}$, below which itinerant quasiparticles screen the impurity spin, is higher than the superconducting transition temperature $T_{\rm c}$, giving rise to the screened-spin ground state known as the local Fermi liquid [Fig.~1(a)]~\cite{Lohneysen_RMP_2007}.
In the case of the impurity spin $S = 1/2$, the ground state is a spin singlet, whereas the lowest excited state, which is known as the Yu-Shiba-Rusinov (YSR) state~\cite{Yu_APS_1965,Shiba_PTP_1968,Rusinov_JETP_1969}, is a spin doublet.
The spin-doublet excited state is realized if the many-body system loses (acquires) an electron that is spin parallel (antiparallel) to the impurity spin.
Thus, the YSR state yields a pair of spin-polarized peaks in the single-particle excitation spectrum at symmetric energies $\pm E_{\rm B}$ within the superconducting gap $\Delta$ [Fig.~1(b)].
In the opposite case, $T_{\rm K} < T_{\rm c}$, the ground state is a spin doublet because no quasiparticles are available to screen the impurity spin [Fig.~1(c)].
The lowest excited state (namely the YSR state) is a spin singlet where the quasiparticles at $|E_{\rm B}|$ screen the impurity spin.
With varying $J$ and thus $T_{\rm K}$, $|E_{\rm{B}}|$ decreases smoothly and crosses zero at $T_{\rm K} \sim T_{\rm c}$, signifying a quantum phase transition between the screened-spin and free-spin ground states. As $J$ increases further inside the screened-spin regime, $|E_{\rm{B}}|$ turns to increase~\cite{Matsuura_PTP_1977,Satori_JPSJ_1992,Sakai_JPSJ_1993,Yoshioka_JPSJ_2000}.
In the case of $S > 1/2$, the impurity spin may not be fully screened, but the same quantum phase transition occurs~\cite{Zitko_PRB_2011,Oppen_PRB_2021}.

The recent development of scanning tunneling microscopy/spectroscopy (STM/STS) has enabled us to investigate magnetic impurities in a superconductor down to the atomic scale, making clear microscopic aspects of the YSR states~\cite{Heinrich_PSS_2018}.
The outcomes have triggered potential applications, including a building block of bottom-up platforms of Majorana quasiparticles~\cite{Perge_Science_2014,Jeon_Science_2017,Ferdman_NP_2016,Kim_SAdv_2018}, functional scanning probes~\cite{Huang_NP_2020, Schneider_SAdv_2021, Huang_PRR_2021}, and a novel quantum bit~\cite{Mishra_PRX_2021}.
It is often crucial to identify which ground state, screened spin or free spin, is realized.
Although the detection of the YSR states by STM/STS is straightforward, it is surprisingly challenging to distinguish the two ground states from the observed tunneling spectrum.
The reason is that the tunneling spectra are qualitatively indistinguishable between the screened-spin and free-spin ground states [Fig.~1(b) and 1(d)].
Various attempts have been made~\cite{Franke_Science_2011,Farinacci_PRL_2018,Hatter_NC_2017,Huang_CP_2020,Kamlapure_NanoLett_2021,Odobesko_QPT_PRB_2020,Karan_NP_2022} by effectively controlling $J$. Since the $J$ control often demands large perturbations to the magnetic impurities, it may be difficult to implement in some applications.

Here we propose a simple yet powerful method based on the Zeeman effect.
The basic idea is depicted in Fig.~1.
We assume the impurity spin $S = 1/2$, but the same argument also applies for $S > 1/2$~\cite{Zitko_PRB_2011,Greven_PRB_2017} (See Appendix A for further details).
In the case of the screened spin ($T_{\rm K} > T_{\rm c}$), applied magnetic field $B$ splits the doublet excited state by the Zeeman effect, whereas the singlet ground state remains intact [Fig.~1(a)].
Therefore, the YSR peaks in the tunneling spectrum show Zeeman splitting [Fig.~1(b)].
If $T_{\rm K} < T_{\rm c}$, the situation is reversed.
The ground state is a spin doublet that exhibits Zeeman splitting, and the excited state is a field-insensitive spin singlet [Fig.~1(c)].
In tunneling spectroscopy, the transition between the Zeeman-split ground states is forbidden because it accompanies a change in the $z$ component of the total spin $S_{\rm{z}}$ by 1, which is incompatible with the injection or extraction of a single electron with a spin $s = 1/2$.
Therefore the lowest detectable excited state remains the singlet state.
If the experiment is performed at a low temperature well below the Zeeman-splitting energy, only the transition from the field-lowered ground state contributes to the YSR peaks in the spectrum, because the field-lifted counterpart is empty.
As a result, the YSR peaks should exhibit a Zeeman shift rather than the Zeeman splitting [Fig.~1(d)].
Consequently, the nature of the ground state of the magnetic impurity in a superconductor should manifest itself in the tunneling spectrum in a magnetic field as a splitting or a shift of the YSR peak.

The ground-state-sensitive Zeeman effect has been observed in the Andreev bound states formed in a quantum dot attached to a superconductor, a formally equivalent system with the magnetic impurity in a superconductor~\cite{Lee_NN_2014,Jellinggaard_PRB_2016}.
Identifying the ground state of the actual magnetic impurity may lead to a breakthrough in spin-related applications, particularly functional scanning probes, because the YSR state is fully spin-polarized in a magnetic field.
In the case of the screened-spin ground state, the pair of Zeeman-split YSR peaks at $|E_{\rm B1}| < |E_{\rm{B}}(B=0)|$ represents the extraction of spin-up electrons from the filled state and the injection of spin-down electrons into the empty state [Fig.~1(a),(b)].
The spin directions are reversed for the pair at $|E_{\rm B2}| > |E_{\rm{B}}(B=0)|$.
(We assume that the magnetic field is applied upward and use the notation where the spin direction is antiparallel to the magnetic moment.)
In the case of the free-spin ground state, only the spin-down (-up) electron can tunnel from (into) the YSR state in the filled (empty) state [Fig.~1(c) and 1(d)].
Besides its perfect spin polarization, the YSR peak is isolated from the quasiparticle continuum above $\Delta$ and thus intrinsically narrow.
These features work in favor of spin-polarized STM/STS.
The unprecedented high spin resolution has been demonstrated~\cite{Schneider_SAdv_2021}, but the quantitative estimation of the spin polarization has remained challenging because the exact ground state of the impurity has been unknown.

Our Zeeman-effect-based method enables systematic identification of the ground state.
The challenge is that the expected Zeeman energy $E_{\rm Z} = g\mu_{\rm B}Bs$, where $\mu_{\rm B}$ is the Bohr magneton, is only 58~$\mu$eV/T if the $g$-factor is 2, demanding a high energy resolution in spectroscopy.
We utilized an ultra-low temperature STM with an energy resolution as high as 20~$\mu$eV to overcome the energy-resolution issue~\cite{Machida_RSI_2018}.
Another problem is that the host superconductivity should be robust even at a high magnetic field where the Zeeman effect becomes detectable.
To this end, we adopted the so-called YSR tip where the YSR states are formed by decorating the apex of a sharp superconducting tip with a single magnetic atom \cite{Huang_NP_2020,Schneider_SAdv_2021}.
The confinement effect suppresses the orbital pair breaking, allowing us to maintain superconductivity well above the bulk upper critical field.

We performed STM/STS experiments using the home-built ultra-high-vacuum (UHV) ultra-low-temperature STM at its base temperature of $\sim 90$~mK~\cite{Machida_RSI_2018}.
A bias voltage $V$ was applied to the Cu(111) surface using the STM controller (Nanonis BP5 and SC5) and an additional 1/100 voltage divider.
A tunneling current $I$ was detected by a current-voltage converter (Femto LCA-1K-5G) that was connected to the Nb tip.
The tunneling spectra were taken by the numerical differentiation of the $I-V$ curves.
Since we are interested in the electronic state of the tip, we flip the sign of $V$ when converting it to the energy, so that the positive energy on the spectra corresponds to the empty state of the tip.

We prepared the superconducting tips by mechanically cutting a polycrystalline Nb wire (3N purity, 0.25-mm diameter) in air.
The apex of the superconducting Nb tip was cleaned by argon-ion sputtering in the UHV chamber. 
To evaluate the robustness of superconductivity in a magnetic field, we measured the tunneling spectra of the Nb tip on the Cu(111) surface, which was prepared by repeating argon-ion sputtering of a Cu single crystal (5N purity) and subsequent annealing at about 550~$^{\circ}$C. 
We chose a tip that exhibited a clear superconducting gap up to $B = 2$~T [Fig.~2(a)].
We formed the YSR states at the apex following the procedure used in Ref. \citenum{Schneider_SAdv_2021}.
We first deposited a small amount of isolated Fe adatoms on the clean Cu(111) surface. 
The deposition was done at about 10 K to avoid the aggregation of the deposited Fe atoms. 
To pick up the Fe adatom, we first stabilized the Nb tip over an isolated single Fe adatom at $V = +20$~mV and $I = 100$~pA.
We then opened the feedback loop and advanced the tip toward the Fe adatom by $\sim 5$~\AA, followed by applying a bias-voltage pulse (+2~V/100~ms).
Immediately after the voltage pulsing, the tip was retracted and the feedback loop was engaged.
The success of the process was confirmed by scanning the same area before and after the pick-up [Fig.~2(b) and 2(c)].
To release the adsorbed Fe atom on the tip, we used the same process with the opposite pulse-voltage polarity.

Figure~3(a) and 3(b) shows the tunneling spectra of two representative tips.
The YSR states are observed as a pair of peaks at $\pm E_{\rm B}$ within the superconducting gap.
The intensity of the YSR peaks and $E_{\rm B}$ often vary if we repeat the dropping and picking up of the Fe atom, indicating that $J$ and $T_{\rm K}$ change from tip to tip.
Sometimes there appear multiple YSR states in the spectrum, suggesting the contributions from multiple orbitals~\cite{Heinrich_PSS_2018}, the effect of magnetic anisotropy~\cite{Heinrich_PSS_2018,Zitko_PRB_2011,Oppen_PRB_2021}, and the multi-channel Kondo screening~\cite{Zitko_PRB_2011,Oppen_PRB_2021}.
To avoid these complications, we selected seven tips where the single pair of YSR peaks dominates the spectrum.
We labeled the tips as \#1$\sim$\#7 in the ascending order of $E_{\rm B}$.
(See Fig.~9 and 10 for the spectra of all the tips.)
We will show below that two tips \#2 and \#3 shown in Figure~3(a) and 3(b), respectively, have different ground states even though they have similar values of $E_{\rm B}$.
The widths of the observed YSR peaks are governed by the thermal broadening, meaning that the YSR states are isolated from non-thermal perturbations.
(See Appendix B.)

Figure~3(c) and 3(d) shows the magnetic-field dependence of the YSR peaks.
Each YSR peak in \#2 splits into two peaks with increasing magnetic field, signifying that the ground state is the screened-spin state.
By contrast, the YSR peaks in \#3 do not split but merely shift, being consistent with the free-spin ground state.
Thus, we have successfully demonstrated the ground-state-sensitive Zeeman effect of the magnetic impurity in a superconductor.

We repeated the magnetic-field dependence measurements in all the tips and found that the tips \#1 and \#3 possess the free-spin ground state, whereas all other tips have the screened-spin ground state.
Figure~4 shows the magnetic-field dependence of the YSR-peak energy.
(See Appendix B for the fitting procedure used to determine the peak energy.)
There is a systematic trend that the YSR-peak energy saturates or even decreases with increasing magnetic field as $|E_{\rm{B}}(B = 0)|$ becomes larger.
This nonlinear field dependence is more pronounced in the tips with the free-spin ground state than those with the screened-spin ground states.
Even though $E_{\rm{B}}(B)$ shows a nonlinearity, the Zeeman-splitting widths $E_{\rm Z}(B)$ for the screened-spin ground state are reasonably linear in $B$, as shown in Fig.~4(h).
The $g$-factors evaluated from the slopes of $E_{\rm Z}(B)$ are smaller than the free-electron value of 2, especially at large $|E_{\rm{B}}(B = 0)|$ [Fig.~4(i)].
We speculate that the interaction between the YSR state and the field-induced quasiparticle continuum near $\Delta$ may be responsible for the nonlinearity in $E_{\rm{B}}(B)$ and the apparently small $g$-factor, but the details are to be investigated.

Next, we investigate the spin polarization of the YSR state by means of spin-dependent tunneling using the configurations illustrated in Fig.~5(a) and 5(b).
The difference in the YSR-peak intensities between the spectra taken on the non-magnetic bare Cu(111) surface [Fig.~5(a)] and the magnetic Fe adatom [Fig.~5(b)] provides us with information on the spin polarization of both Fe atoms.
An isolated Fe adatom on Cu(111) carries a magnetic moment $\sim 3.5 \mu_{\rm B}$ on average, and its single-atom magnetization curve saturates above $B \sim 0.2$~T~\cite{Khajetoorians_PRL_2011}.
A similar situation may occur on the Nb tip~\cite{Odobesko_PRB_2020}, and thus the magnetic moments of the Fe atoms on the tip and on the Cu(111) surface should be aligned parallel to the magnetic field in the field range we used, $B \geq 0.5$~T.
To avoid the influence of the atom-to-atom variations of the Fe magnetic moment on Cu(111)~\cite{Khajetoorians_PRL_2011}, we performed all the measurements on the same Fe adatom shown in Fig.~5(c).

Figure~5(d)-5(g) compares the tunneling spectra taken on the bare Cu(111) surface and the Fe adatom using the tips \#2 and \#3, which have the screened- and free-spin ground states, respectively.
In both cases, the intensities of the field-lowered peaks are enhanced on the Fe adatom, whereas those of the field-lifted peaks are suppressed compared to their counterparts on the bare Cu(111) surface.
These observations evidence the spin polarization of the YSR states.
Since the peak enhanced on the Fe adatom is the field-lowered one, which is spin-down, the dominant spin component of the Fe-adatom state on Cu(111) surface at the Fermi level should also be spin-down.
This result apparently contradicts the result of the first-principles calculations where the minority-spin component being spin-up dominates the density-of-states near the Fermi level~\cite{Lazarovits_PRB_2003}.
A similar discrepancy has also been reported in the ferromagnet/insulator/superconductor planar tunnel junctions, where the spin-polarizations of the ferromagnets are estimated using the Zeeman splitting of the coherence peak in the superconducting-gap spectrum~\cite{Meservey_PhysRep_1994,Zutic_RMP_2004}.
In this case, it has been argued that the tunneling matrix element may be orbital and momentum dependent, causing the difference from the orbital- and momentum-integrated first-principles-calculations result.
It is an important future issue whether the matrix element also matters for the YSR-state-based spin-polarized STM, where the orbital and momentum selectivity should be different from the planar tunneling case.

Finally, we evaluate the spin polarization $P$ defined using the normalized asymmetry.

\begin{equation}
P \equiv \frac{(W_-^{\rm Fe}/W_-^{\rm Cu})-(W_+^{\rm Fe}/W_+^{\rm Cu})}{(W_-^{\rm Fe}/W_-^{\rm Cu})+(W_+^{\rm Fe}/W_+^{\rm Cu})},
\end{equation}

\noindent
where $W$ is the weight of the YSR peak, the superscripts denote the tunneling locations, and the subscripts $-$ and $+$ indicate the field-lowered and field-lifted YSR peaks, respectively.
(See Appendix B for the fitting procedure used to estimate the peak weights.)
We evaluated $P$ using different tips and different pairs of the YSR peaks.
As shown in Fig.~5(h), $P$ takes a similar value of $\sim 0.6$ at high fields.
Given the perfect spin polarization and the narrow line width ($\lesssim$~100~$\mu$eV)  of the YSR state, this value should represent the spin polarization of the Fe adatom on Cu(111) at the Fermi level.

The Zeeman-effect-based identification of the ground state of the magnetic impurity in a superconductor and the successful demonstration of the YSR-state-based spin-polarized STM should yield various immediate applications, such as direct imaging of the spin structure expected in the Majorana bound state in the vortex core~\cite{Nagai_JPSJ_2014}.
A superconducting tip is not the only system to observe the Zeeman effect of the YSR state.
A thin-film superconductor in a parallel field offers another promising platform, enabling a more controllable manipulation of the magnetic atoms on the surface.
We anticipate that our observations pave the way for these potential applications.

\section*{Acknowledgments}
The authors thank C. J. Butler, Y. Kohsaka, N. Nagaosa, and R. \v{Z}itko for their valuable comments.
This work was supported by JST PRESTO No. JPMJPR19L8, JST CREST No. JPMJCR16F2, and Grants-in-Aid for Scientific Research (KAKENHI) 19H01843.



\appendix
\section{Zeeman effects on the YSR states for $\textbf{\textit{S}}$$\mathbf{~>~1/2}$}
In the main text, we argued the case of the bare impurity spin $S = 1/2$ where the YSR peak in the tunneling spectrum splits or shifts in a magnetic field if the many-body ground state is screened or free, respectively.
Here we show that the same behavior is expected even in the case of $S > 1/2$.
In the following, we will discuss the case of $S = 1$, but the same argument applies for any $S$.
Because we focus on the situation where a single pair of the YSR states dominate the spectrum, we neglect the effects of magnetic anisotropy and multi-channel Kondo screening~\cite{Heinrich_PSS_2018,Zitko_PRB_2011,Oppen_PRB_2021}.

If $S > 1/2$, the impurity spin cannot be fully screened even in the case of the screened-spin ground state $T_{\rm K}~>~T_{\rm c}$.
In the case of $S = 1$, the screened-spin ground state and the free-spin excited state have total spins $S = 1/2$ and $S = 1$, respectively [Fig.~6(a)].
Therefore, in a magnetic field, the Zeeman effect splits both states into sub-states with different $S_z$, a $z$-component of $S$. 
Considering that only the lowest-energy sub-state is populated at a sufficiently low temperature and only the pair of sub-states with $S_z$'s differ by 1/2 participates the tunneling spectrum, the observable transitions are from $S_z = -1/2$ to $S_z = -1$ and $S_z = 0$.
As a result, the Zeeman splitting of the YSR peak is expected, similar to the case of $S~=~1/2$.

In the case of the free-spin ground state $T_{\rm K}~<~T_{\rm c}$, the ground state and the lowest excited state possess $S = 1$ and $S = 1/2$, respectively [Fig.~6(b)].
Since the only possible transition is from $S_z = -1$ to $S_z = -1/2$, the YSR peak should not split but merely shift in a field, again similar to the case $S = 1/2$.

\section{Fitting the YSR peaks}
We fitted the tunneling spectra to extract the peak energies and the weights of the observed YSR peaks.
We adopted a Voigt function to describe each YSR peak.
A density-of-states spectrum associated with the quasiparticle continuum, which overlaps with the YSR peaks at a high magnetic field, was represented by a Dynes function with a phenomenological broadening parameter and an offset.
We show typical fitting results for the tip \#2 at $B = 1.5$~T in Fig.~7 where the definitions of the YSR-peak weights are depicted.

To gain an insight into the factors that determine the width of the YSR peak, we examined the shape of the YSR peak at zero magnetic field.
As shown in Fig.~8, the Lorentzian function, a general form of the damped resonance, cannot reproduce the observed YSR peaks.
Rather, they are well fitted with the energy derivative of the Fermi-Dirac function (d$f(E)$/d$E$) with an effective temperature of 119~$\sim$130~mK (Fig.~8).
These values are reasonably close to the lowest electron temperature in our system ($\sim 90$~mK) estimated by fitting the superconducting-gap spectrum of aluminum~\cite{Machida_RSI_2018}, suggesting that the thermal broadening dominates the width of the YSR peak.
We also fitted the YSR peaks with the Lorentzian functions convoluted with a rectangular function and an energy derivative of the Fermi-Dirac function, which represent the resolution function of the numerical differentiation and the thermal broadening, respectively.
The fitting results provide negligibly small Lorentzian widths ($< 1$~$\mu$eV), confirming that the non-thermal perturbations hardly broaden the YSR peaks.
Such a narrow intrinsic width of the YSR peak is consistent with the result of the so-called Shiba-Shiba tunneling measurements~\cite{Huang_NP_2020}.

Each YSR peak gets broadened with increasing magnetic field, indicating that additional damping factors, such as an interaction with the field-induced low-energy itinerant quasiparticles, cause the lifetime broadening.
To phenomenologically take into account the multiple factors that determine the width of the YSR peak, we adopted a Voigt function, a convolution of the Lorentzian and Gaussian functions, as a fitting function to obtain the energy and the weight of the YSR peak.
As shown in Fig.~9 and 10, the fitting with multiple Voigt functions and the superconducting-gap background reasonably works in the whole field range.

\bibliography{YSR_tip_Refs}

\clearpage
\begin{figure*}[t]
    \begin{center}
    \includegraphics[width=16cm]{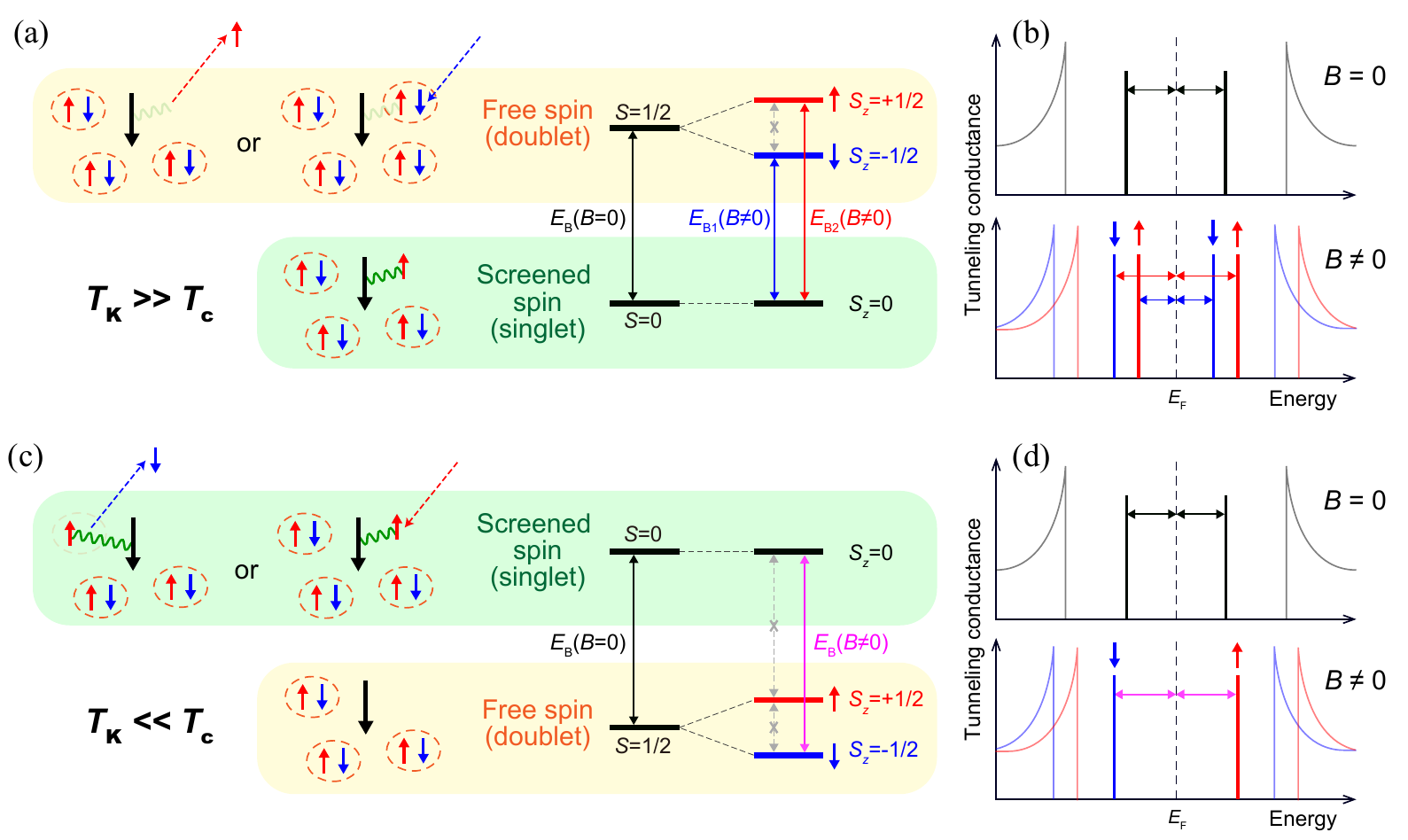}
    \end{center}
    \caption{
    \setlength{\baselineskip}{14pt}
    (a) Schematic diagrams of the spin configurations and many-body-state energy diagrams of the magnetic impurity in a superconductor with the screened-spin ground state.
    A black arrow represents the impurity spin.
    Red and blue arrows denote spins of itinerant electrons that form a Cooper pair represented by the dashed ellipse.
    A green wavy line indicates the antiferromagnetic coupling that causes the Kondo screening.
    The same energy diagram is obtained if all the spin directions are reversed at $B = 0$.
    However, if $B \neq 0$ (on the right of the energy diagram), the states with the impurity spins parallel and anti-parallel to $B$ are no longer degenerate, giving rise to the Zeeman splitting.
    (b) Expected tunneling spectra at $B = 0$ and $B \neq 0$ in the case of the screened-spin ground state.
    Double-headed arrows correspond to $E_{\rm B}$ in (a) denoted by the same colors.
    (c) and (d) The Same as (a) and (b), respectively, but for the free-spin ground state ($T_{\rm K} < T_{\rm c}$).
    }
    \end{figure*}
    
    \begin{figure*}
        \begin{center}
        \includegraphics[width=16cm]{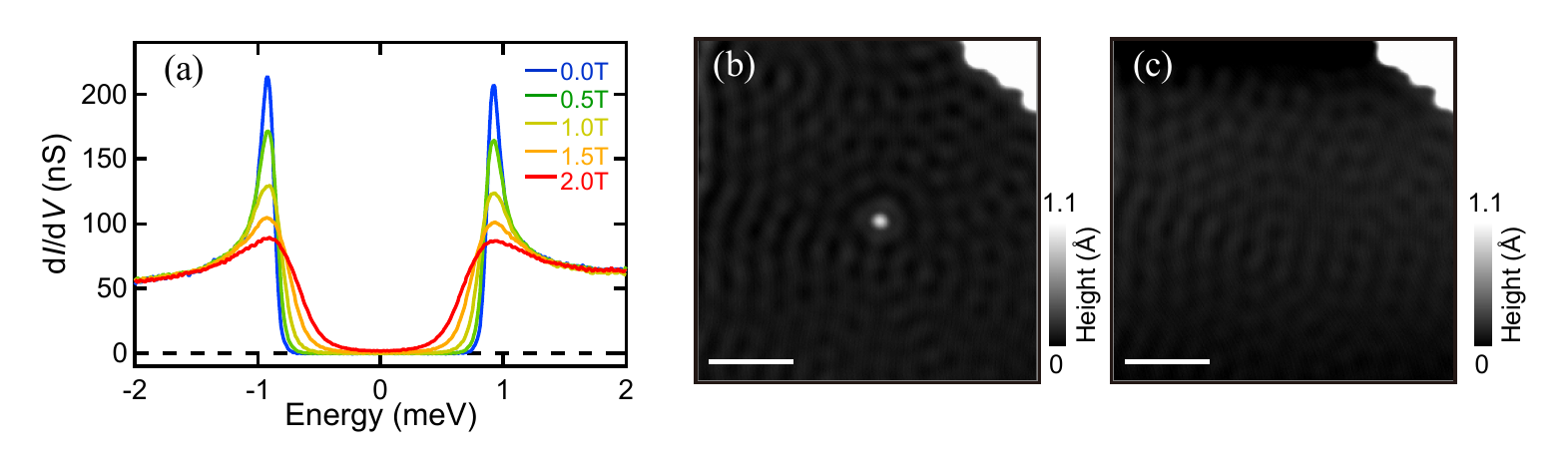}
        \end{center}
        \caption{
        \setlength{\baselineskip}{14pt}
        (a) Magnetic-field dependence of the tunneling spectra taken between the clean Cu(111) surface and a Nb tip without an Fe atom.
        The junction was stabilized at $V = +20$~mV and $I = 1$~nA before measuring the $I-V$ curves.
        Magnetic fields were applied perpendicular to the Cu(111) surface.
        The superconducting gap survives above the bulk upper critical field of 0.42~T.
        (b) and (c) Two subsequent STM topographic images before and after the picking-up process of the Fe adatom.
        Imaging conditions are $V = +20$~mV and $I = 10$~pA.
        Scale bars correspond to 5~nm.
        }
    \end{figure*}
    
    \begin{figure*}[t]
        \begin{center}
        \includegraphics[width=16cm]{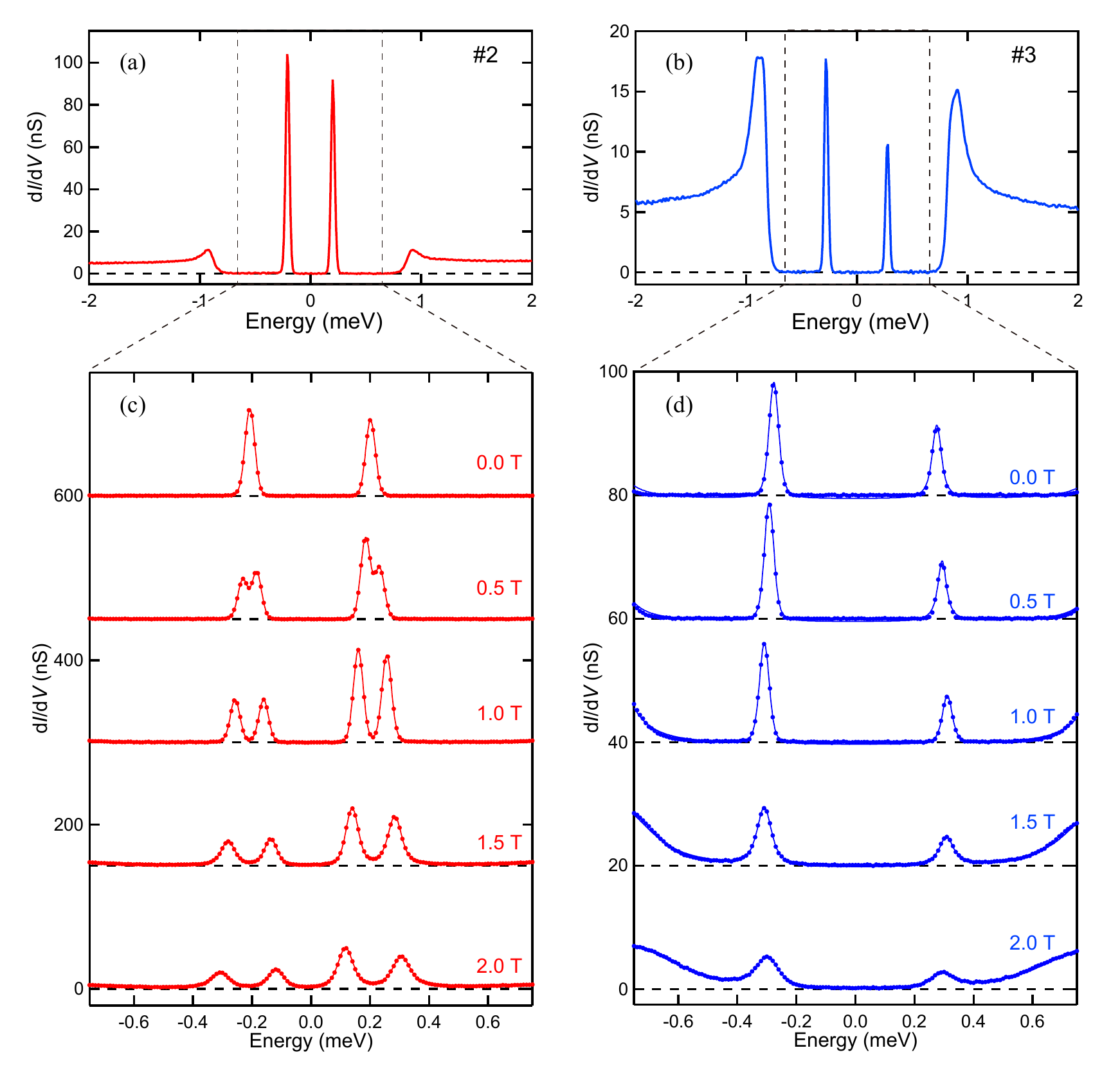}
        \end{center}
        \caption{
        \setlength{\baselineskip}{14pt}
        (a) A tunneling conductance $dI/dV$ spectrum of the tip \#2 showing the YSR peaks at $E_{\rm B} = \pm$207~$\mu$eV within the superconducting gap $\Delta \sim 900$~$\mu$eV.
        (b) A $dI/dV$ spectrum of the tip \#3 where the YSR peaks appear at $E_{\rm B} = \pm$278~$\mu$eV.
        (c) Magnetic-field dependence of the YSR peaks in \#2.
        Each YSR peak shows Zeeman splitting, signifying the screened-spin ground state.
        (d) Magnetic-field dependence of the YSR peaks in \#3.
        Each YSR peak shifts to higher $|E|$ but does not split, indicating the free-spin ground state.
        In (c) and (d), the filled circles denote data points and the solid lines show the results of the multi-peak fitting.
        (See Appendix B for the multi-peak fitting.)
        Each spectrum is offset vertically for clarity.
        We stabilized the tip at a bias voltage to the Cu(111) surface $V= +20$~mV and at a tunneling current $I = 100$~pA.}
    \end{figure*}

    \begin{figure*}[t]
        \begin{center}
        \includegraphics[width=15cm]{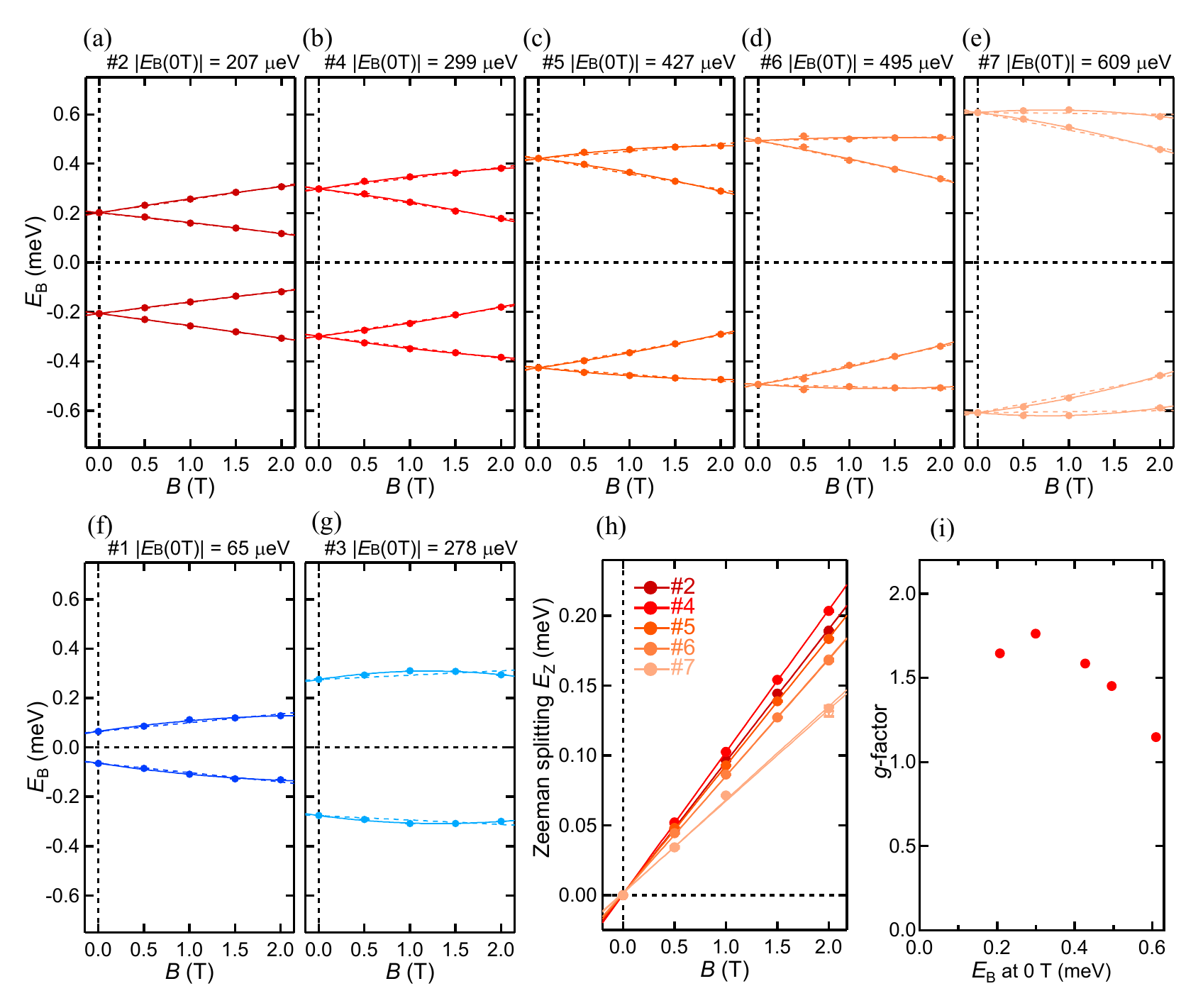}
        \end{center}
        \caption{
        \setlength{\baselineskip}{14pt}
        (a-e) Magnetic field dependence of the YSR-peak energies $E_{\rm{B}}$ for the tips that have a screened-spin ground state.
        (f) and (g) Magnetic field dependence of the YSR-peak energies for the tips that have a free-spin ground state.
        The peak energies are determined by the multi-peak fitting.
        (See Appendix B for the multi-peak fitting.)
        The dashed and solid lines denote the results of the fitting of the peak energies to the $B$-polynomials up to $B$ (straight line) and $B^2$, respectively.
        (h) The Zeeman-splitting widths $E_{\rm{Z}}$ for the tips with the screened-spin ground state.
        The solid lines result from the straight-line fitting, providing the $g$-factors as slopes.
        (i) The $g$-factors as a function of the YSR-peak energy at $B =0$~T.
        }
    \end{figure*}
    
    \begin{figure*}[t]
        \begin{center}
        \includegraphics[width=15cm]{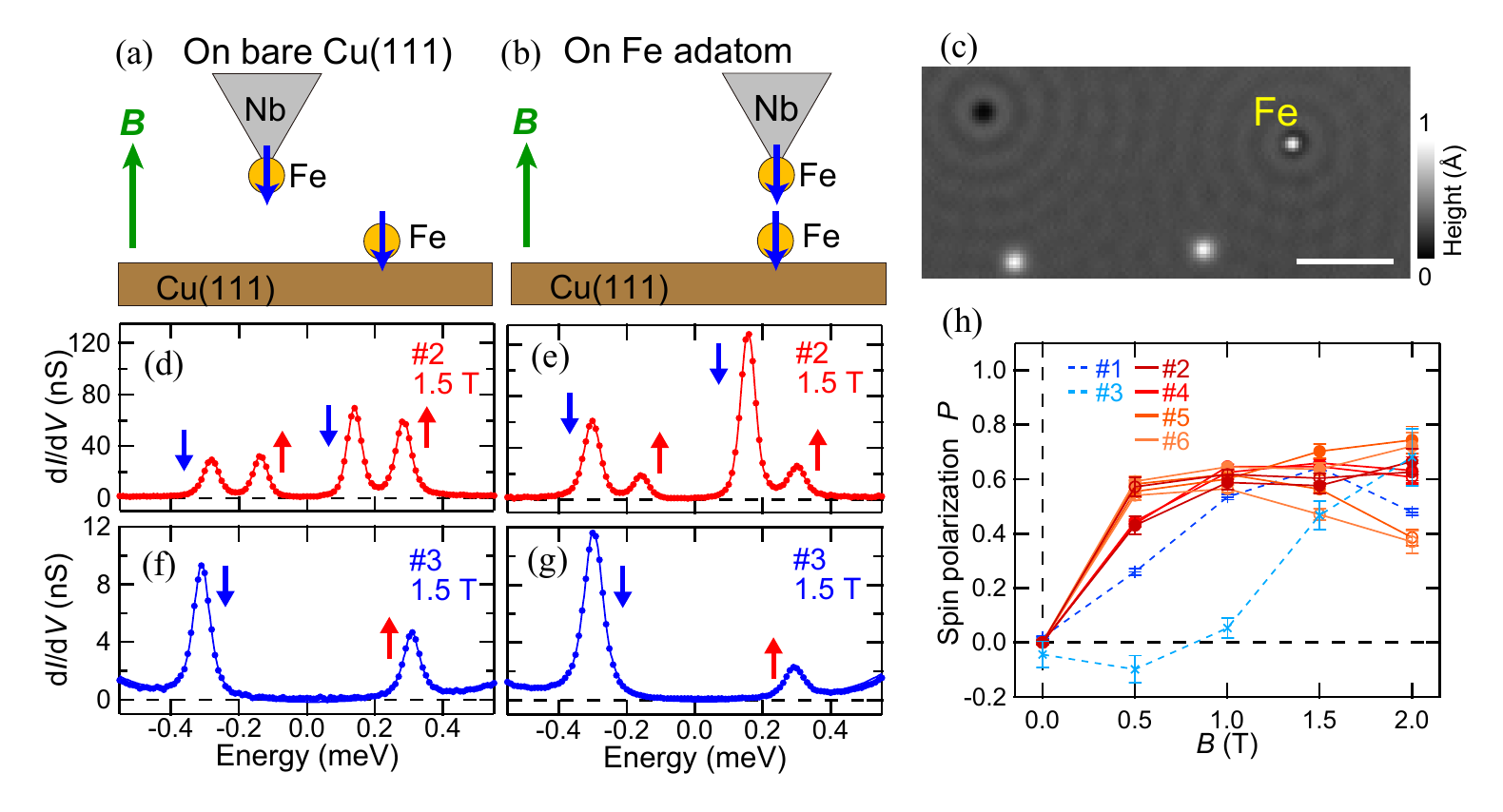}
        \end{center}
        \caption{
        \setlength{\baselineskip}{14pt}
        (a) and (b) Schematic illustrations of the experimental configurations used for the YSR-state-based spin-polarized STM/STS measurements. Arrows on the Fe atoms represent the directions of the majority spin of the Fe atoms, which are antiparallel to the magnetic moments.
        (c) An STM topographic image near the Fe adatom on which the spectroscopic measurements have been carried out.
        Reference spectra on the bare Cu(111) surface were taken at clean locations at least 5~nm away from the Fe adatom.
        A depression at the top left corner and two protrusions at the bottom are non-magnetic native defects on the Cu(111) surface.
        Imaging conditions are $V = +20$~mV and $I = 10$~pA.
        The scale bar denotes 5~nm.
        (d) and (e) Tunneling spectra of the tip \#2 (screened-spin ground state) on the bare Cu(111) surface and on the Fe adatom, respectively.
        (f) and (g) Tunneling spectra of the tip \#3 (free-spin ground state) on the bare Cu(111) surface and on the Fe adatom, respectively.
        In (d-g), the spin-down and spin-up YSR states are enhanced and suppressed, respectively, on the Fe adatom.
        The tip stabilization conditions are the same as in Fig.~3.
        Magnetic field of 1.5~T was applied perpendicular to the Cu(111) surface.
        (h) Magnetic-field dependence of the spin polarizations estimated using various tips and different pairs of the YSR states.
        The solid and dashed lines denote the results of the tips with the screened-spin and free-spin ground states, respectively.
        The filled and open circles represent the data obtained in the tips with the screened-spin ground state using the Zeeman-split YSR states in the filled and empty states, respectively.
        }
    \end{figure*}

    \begin{figure*}[h]
        \begin{center}
        \includegraphics[width=15cm]{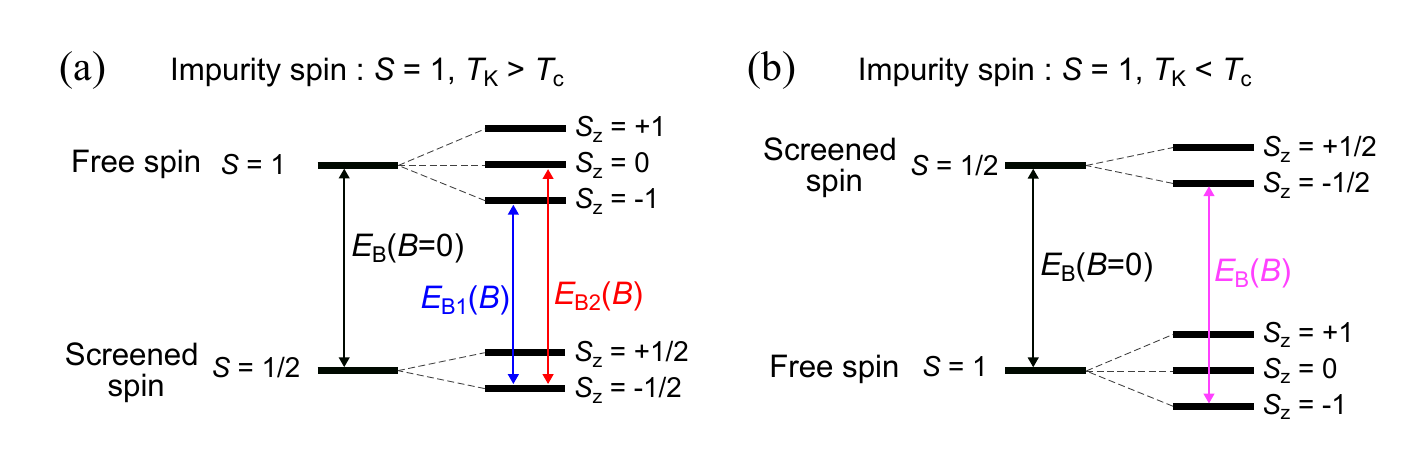}
        \end{center}
        \caption{
        \setlength{\baselineskip}{14pt}
        (a), Many-body-state energy diagrams of the magnetic impurity with $S~=~1$ in a superconductor with the screened-spin ground state ($T_{\rm{K}}~>~T_{\rm{c}}$) at a magnetic field $B~=~0$ (left) and $B~\neq~0$ (right). 
        (b), The same as (a), but for the free-spin ground state ($T_{\rm{K}}~<~T_{\rm{c}}$).
        }
    \end{figure*}
    
    \begin{figure}[h]
    \begin{center}
    \includegraphics[width=12cm]{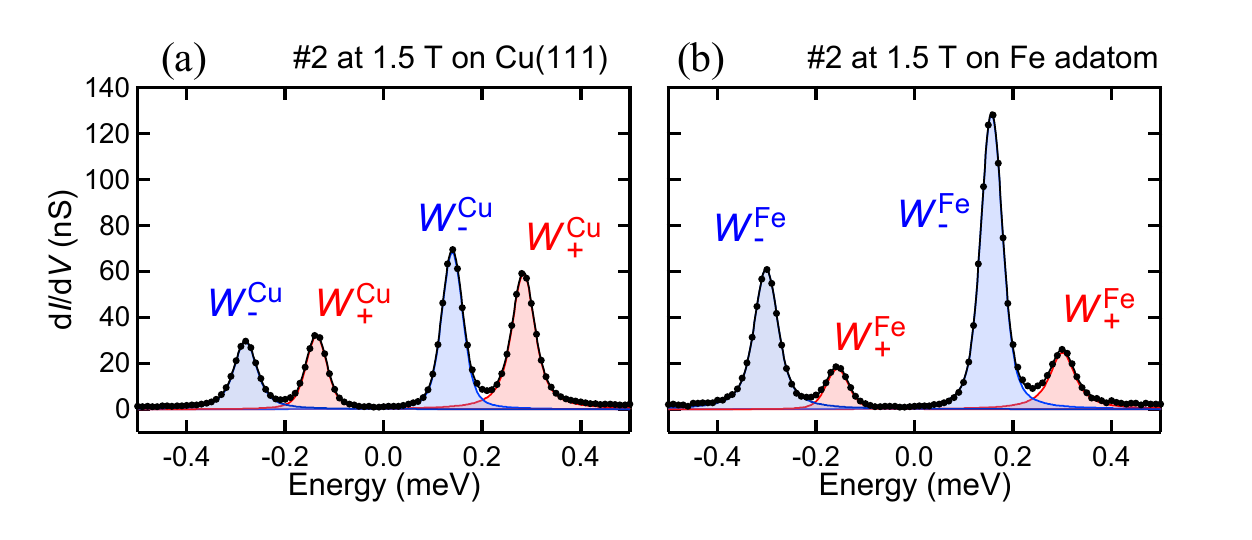}
    \end{center}
    \caption{
    \setlength{\baselineskip}{14pt}
    (a) and (b), Tunneling spectra of tip \#2 at $B = 1.5$~T taken on the Cu(111) surface and on the Fe adatom, respectively.
    Filled black circles and a black line represent the raw data points and the fitting result using the multiple-Voigt functions, respectively.
    Each color-shaded area shows the weight of the YSR-peak component used to estimate the spin polarization shown in Figure~5h.
    }
    \end{figure}
    
    \begin{figure*}[h]
        \begin{center}
        \includegraphics[width=17cm]{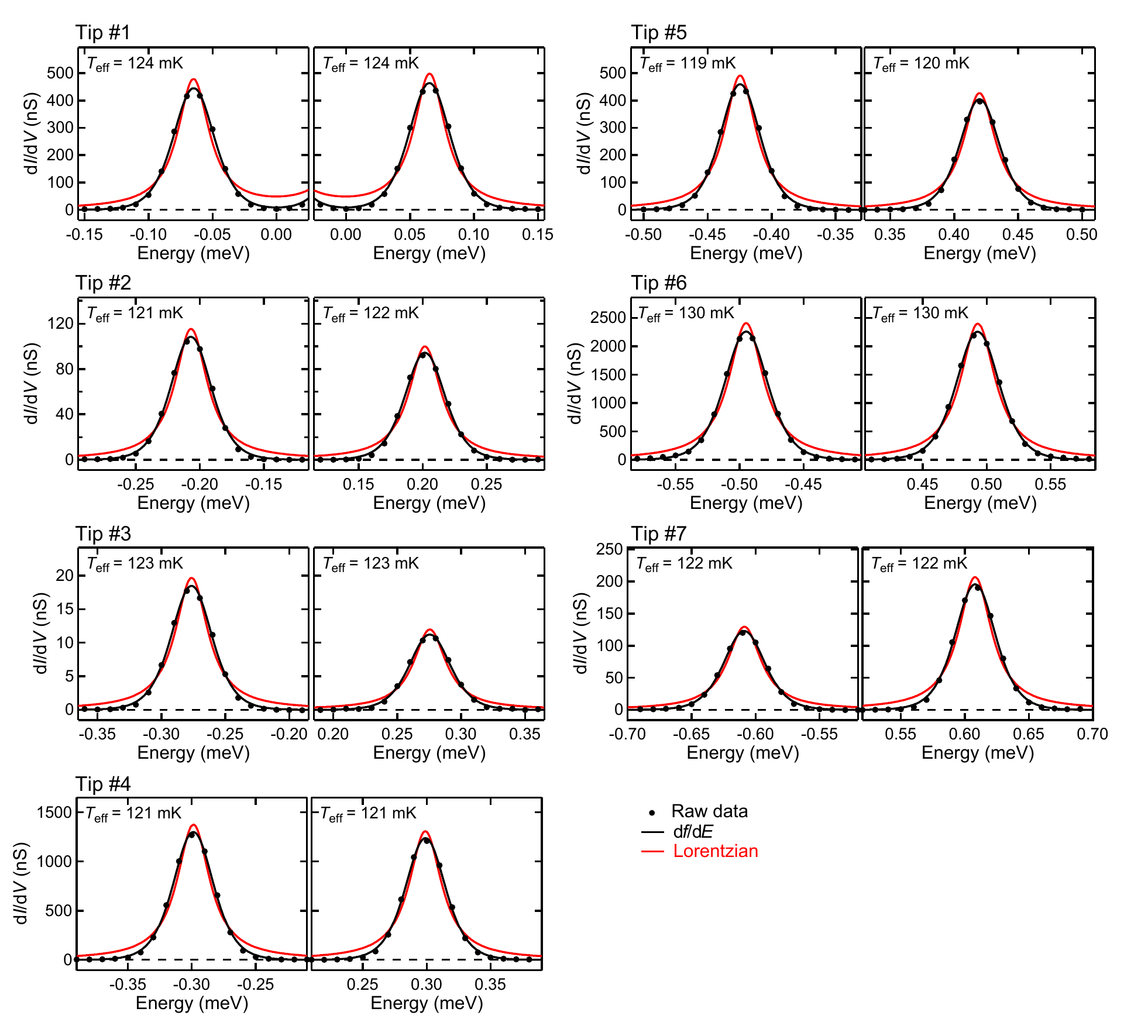}
        \end{center}
        \caption{
        \setlength{\baselineskip}{14pt}
        Tunneling spectra near the YSR peaks.
        Data were taken on the Cu(111) surface at $B~=~0$~T.
        Filled black circles denote the data points.
        Black and red curves denote the results of fitting the energy-derivative of the Fermi-Dirac function and the Lorentzian function, respectively. 
        The effective temperature $T_{\rm{eff}}$ obtained by the fitting to the energy-derivative of the Fermi-Dirac function is shown in each panel.
        }
    \end{figure*}
    
    \begin{figure*}[h]
        \begin{center}
        \includegraphics[width=16.0cm]{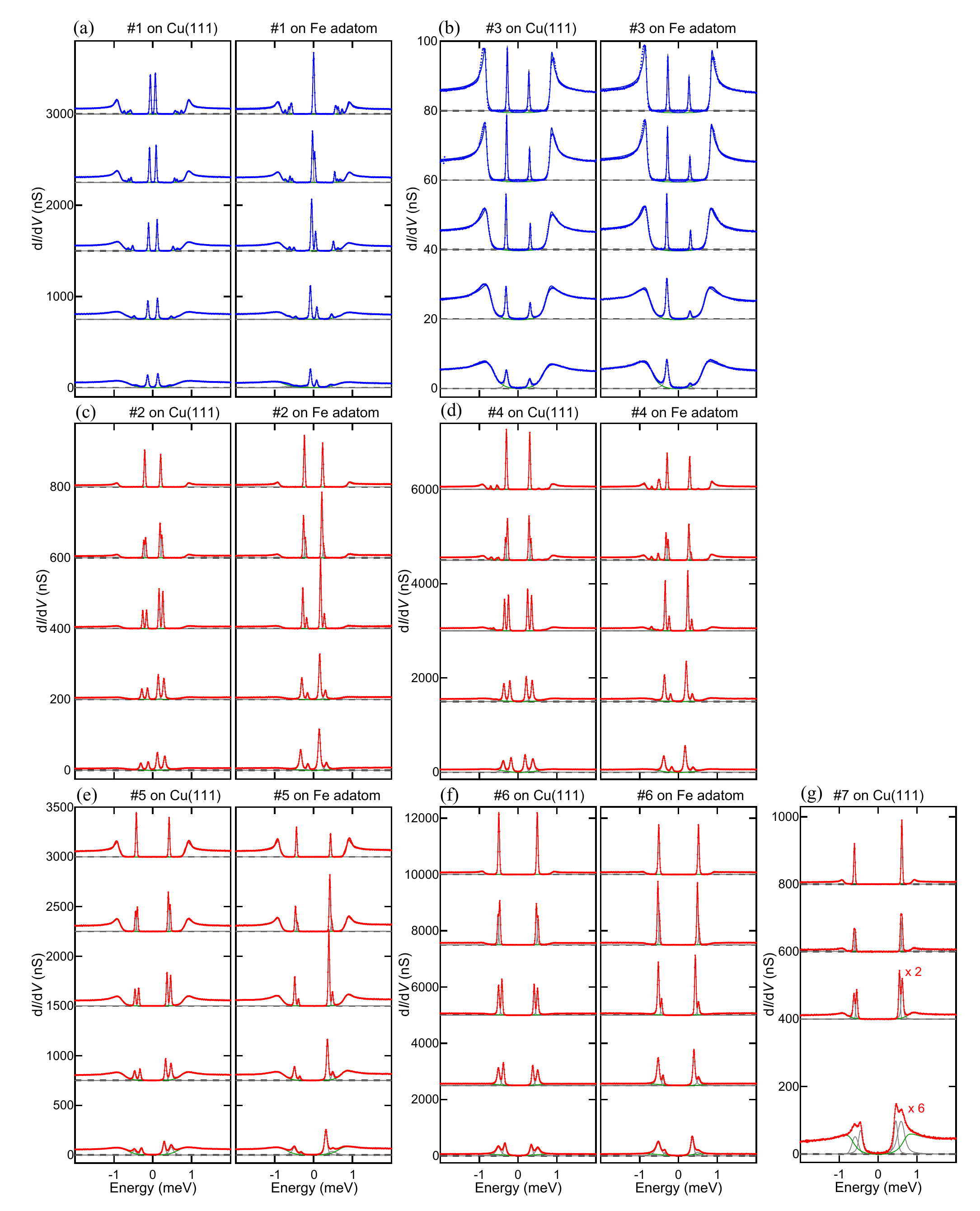}
        \end{center}
        \caption{
        \setlength{\baselineskip}{14pt}
        For each tip, left and right columns show the spectra taken on the bare Cu(111) surface and on the Fe adatom, respectively.
        No data on the Fe adatom were available in the tip \#7 because it became unstable during the measurement.
        Spectra in each column show the data at $B = 0$~T, 0.5~T, 1.0~T, 1.5~T, and 2.0~T from top to bottom.
        Gray lines denote the Voigt functions used to fit the YSR peaks.
        Green lines represent the background density-of-states spectra that we approximate using Dynes functions with phenomenological broadening parameters and offsets.
        The set points are $V~=~$+20~mV and $I~=~100$~pA for \#2 and \#3 and $I~=~1$~nA for others.
        }
    \end{figure*}
    
    \begin{figure*}[h]
        \begin{center}
        \includegraphics[width=16cm]{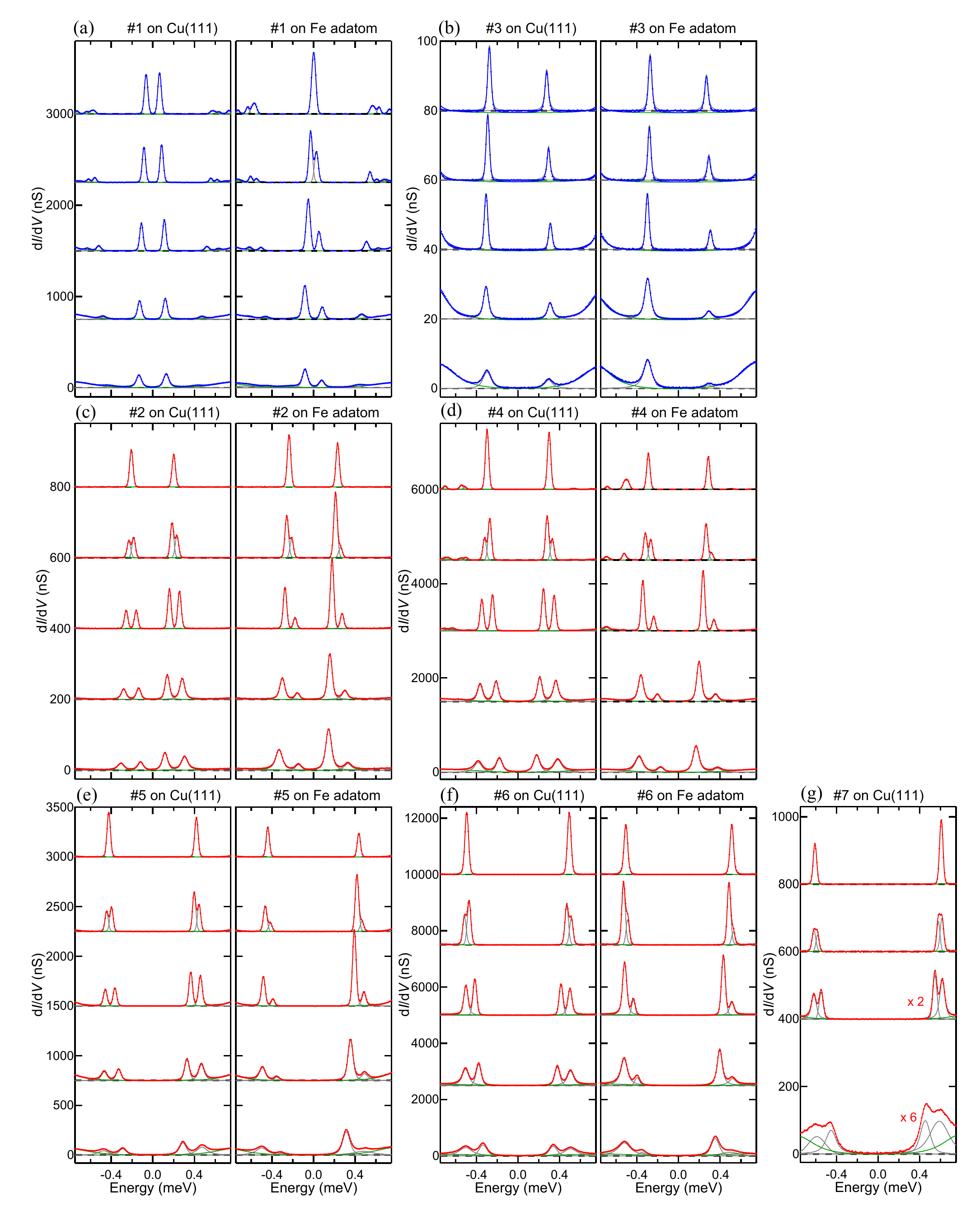}
        \end{center}
        \caption{
        \setlength{\baselineskip}{14pt}
        The same as Fig.~9, but the energy ranges are $\pm 0.75$~meV.
        }
    \end{figure*}
\end{document}